\pdfoutput=1
\documentclass[a4paper,11pt]{article}
\pdfoutput=1 

\usepackage{jcappub} 

\usepackage[T1]{fontenc} 
\usepackage{epsf}
\usepackage{epstopdf}

\title{Cosmological Model Independent Time Delay Method}

\author[a,e]{Yi Zhang,}
\author[b,e]{Xuewen Liu,}
\author[c]{Jingzhao Qi,}
\author[d,e,1]{Hongsheng Zhang,\note{Corresponding author.}}

\affiliation[a]{College of  Science, Chongqing University
of Posts and Telecommunications,  \\Chongqing, 400065, China}
\affiliation[b]{ School of Physical Sciences, University of Chinese Academy of Sciences, \\No.19A Yuquan Road, Beijing 100049, China}
\affiliation[c]{ Department of Astronomy, Beijing Normal University, \\Beijing, 100875, China}
\affiliation[d]{ School of Physics and Technology, University of Jinan, \\West Road of Nan Xinzhuang 336, Jinan, Shandong 250022, China}
\affiliation[e]{ CAS Key Laboratory of Theoretical Physics, Institute of Theoretical Physics, Chinese Academy of Sciences, \\Beijing, 100190, China}

\emailAdd{zhangyia@cqupt.edu.cn}
\emailAdd{ liuxuewen14@itp.ac.cn}
\emailAdd{tqijingzhao@bnu.edu.cn}
\emailAdd{sps\_zhanghs@ujn.edu.cn}

\abstract{We propose   a Cosmological Model Independent Time Delay   (CMITD)   method where the Lorentz invariance violation (LIV) variable $K(z)$   is constructed by observational data instead of cosmological model.    The simulated time delay data show the CMITD method could present the  validity of  LIV test.  And,   the errors in the propagating process is critical for the existence and magnitude  of LIV. }

\begin{document}
\maketitle
\flushbottom

\section{Introduction}\label{sec1}
Lorentz invariance is the cornerstone of modern physics. As  a non-general symmetry,   the  violation of Lorentz symmetry is expected in  quantum gravity frameworks.  And Lorentz invariance violation (LIV)  deduces a deformed energy-momentum relation  in the high energy scale often around Planck scale  \cite{Mattingly:2005re,Liberati:2013xla,AmelinoCamelia:1997gz,Ellis:2011ek}. As  the velocities of photons are changed, it is also  called modified dispersion relation. Then, the simplest test of LIV is  the arrival-time differences of photons in astrophysics. Two photons of different energies in the LIV background   would lead to   different arrival times which is called time delay,  even they emitted simultaneously from the same remote cosmological source.

Theoretically,  LIV is caused by the quantum gravities which have to modify General Relativity (GR).  Meanwhile, the LIV effect  is small  but  could be directly applied to kinematic process luckily. 
Gamma-ray bursts (GRBs) are  suitable for this kinematic studies with the distant transient properties \cite{AmelinoCamelia:2000zs,Piran:2004qe, Pfeifer:2018pty,DeAngelis:2016qob,RodriguezMartinez:2006ee,Vasileiou:2013vra}.  Moreover, the kinematic process  needs to consider the late acceleration in our universe which could be explained as dark energy model or modified gravity model.  Lorentz symmetry are invariant in  the cosmological  model with GR background.  The dark energy models  with GR background  are used for LIV test in literatures \cite{Ellis:2002in,Ellis:2005wr,Zou:2017ksd,Jacob:2008bw,Pan:2015cqa}.    However, the whole LIV test scenario  is  self-contradictory if the cosmological model in GR was used.

  In this paper, we  calculate the time delay term by using observational data instead of cosmological model. We call it Cosmological Model Independent Time Delay (CMITD) method. The paper is arranged as below. In Section \ref{sec2}, the LIV theories  are introduced. In Section \ref{sec3}, we show the contradictions in theories and our CMITD solution. In Section \ref{sec4}, the  results of LIV test are presented and discussed. Finally, we give a short summary in Section \ref{sec5}.

\section{ Lorentz  Invariance Violation Theory}\label{sec2}

In General Relativity,  the trajectory of the massless particle with  dispersion relation  $E=p^{2}c^{2}$ which shows the velocity of photon is  constant and does not have LIV \cite{weinberg72}.  Violations of local Lorentz invariance modify the dispersion relation  \cite{Yunes:2016jcc}. The broken energy scale named $E_{LV}$ is usually assumed around the Planck scale. When examining particles with energies much smaller than the symmetry breaking scale, we may choose only the leading order correction phenomenally.
 Assuming the leading LIV correction is of order $n$, the LIV model  can
be described as
\begin{equation}
\label{disperision}
E^{2}= p^{2}c^{2}\times [1-s_{\pm}(\frac{E}{E_{LV}})^{n}],
\end{equation}
where   $s_{\pm}=+1/s_{\pm}=-1$ corresponds the subluminal/superluminal case \cite{Vasileiou:2013vra}.

Then,  the modified velocity of photon  is given by 
 \begin{eqnarray}
 \label{v}
 v=\frac{\partial E}{\partial p} =c[1-s_{\pm} \frac{n+1}{2}(\frac{E}{E_{LV}})^{n}].
 \end{eqnarray}
And, the  energy varying  velocities   deduce  a time delay \cite{Jacob:2008bw}
 \begin{eqnarray}
 \Delta t_{LV} =  t_{h^{*}} - t_{l^{*}} = s_{\pm}\frac{(1+n)(E_{h^{*}}^{n}-E_{l^{*}}^{n})}{2H_0 E_{LV}^{n}}\int_{0}^{z}\frac{(1+z')^{n}}{h(z')}dz',
 \end{eqnarray}
where  the index $h^{*}$ ($l^{*}$) denotes high (low) energy, $E_{h^{*}}$ and $E_{l^{*}}$ are dependent on experiments, $h(z)=H(z)/H_0$ with $H(z)$ as the Hubble parameter and the index $0$ denotes today's value.
  For latter convenience, we define the LIV parameter $a_{LV}$ and the LIV variable $K(z)$ as
  \begin{eqnarray}
  \label{alv}
  &&a_{LV}= s_{\pm}\frac{(1+n)( E_{h^{*}}^{n}-E_{l^{*}}^{n})}{2H_0E_{LV}^{n}},\\
  \label{K}
 &&K(z)=\int_{0}^{z}\frac{(1+z')^{n}}{h(z')}dz' .
 \end{eqnarray}
 According to Eq.(\ref{v}), we could have
$|1-v/c|\sim a_{LV}H_0$ where $H_0$ is canceled by $a_{LV}$. The value of $|1-v/c|$ and $E_{LV}$ are independent of $H_0$ intrinsically.  And,
in this paper, our definition of   $K(z)$   is different from  the $K(z)$ in Ref.\cite{Ellis:2005wr} by a factor of $1/(1+z)$ for convience. Moreover,
the value of $n$ is based on gravity theories. Specifically,  $n=0.5$  corresponds to the typical choice of Multifractal Spacetime Theory  \cite{Calcagni:2009kc,Calcagni:2011kn,Calcagni:2011sz,Calcagni:2016zqv}  where  the availbale  range  is $0<n<1$.
And the $n=1$ case corresponds to the
Double Special Relativity \cite{AmelinoCamelia:2000ge,Magueijo:2001cr,AmelinoCamelia:2002wr,AmelinoCamelia:2010pd}.  The $n=2$ case corresponds to Extra-Dimensional Theories \cite{Sefiedgar:2010we}  or  Horava-Lifshitz Gravity \cite{Horava:2008ih,Horava:2009uw,Vacaru:2010rd,Blas:2011zd}. 

As for the LIV parameter $n$,  it needs to  emphasize that every value of $n$ is related to LIV.  Only when the LIV breaking scale $E_{LV}$ approaches  $\infty$,  there is  no Lorentz invariance violation which makes $a_{LV}=0$.   Take $n=0$ for example.  By setting $n=0$ and  $s_{\pm}=+1$,  we could get $E^2=0$ which is not physical. And by  setting $n=0$ and $s_{\pm}=-1$,  we could get  $E^2=2p^2c^2$.  The number $2$  shows the  broken Lorentz symmetry effects with velocity of $\sqrt{2c}$ but  without physical background as well.  Therefore, we would not consider the $n=0$ case.  
But when $n=0$, $K(z)$ is the dimensionless proper distance  $D(z)$   which is defined  as
\begin{eqnarray}
 \label{dl}
D(z)=\int_{0}^{z}\frac{dz'}{h(z')}.
\end{eqnarray}
Then $K(z)$ which is determined by $n$ and $h(z)$ could be called   cosmological-distance-like variable.

And, the time delay $\Delta t_{LV}$ induced by Lorentz invariance violation is likely to be accompanied by an intrinsic energy-dependent time delay  from unknown properties of the source, the observed time delay data include two parts:
\begin{equation}
\label{testeq}
\Delta t_{obs} = \Delta t_{LV}+\Delta t_{int}= a_{LV}K+b_{sf}(1+z).
\end{equation}
The slope $a_{LV}$   is connected to the scale of Lorentz violation and the intercept $b_{sf}$ represents the possible unknown intrinsic time-lag inherited from the source.

\section{ Cosmological Model Independent Time Delay Method}\label{sec3}
General Relativity  has no LIV.  Putting any model related to GR  into the LIV test  is a  wrong assumption. 
Detailedly, if the  $h(z)$ part in Eq.(\ref{K}) is derived from a certain cosmological model in GR, the reduced $K(z)$ is not suitable to constrain the LIV parameter $a_{LV}$. The non-LIV  assumption makes $a_{LV}=0$.  And as $a_{LV}$ is multiplied by $K(z)$ in  Eq.(\ref{testeq}), it   wipes out  $K(z)$'s value. Then, the value of $K(z)$  is meaningless in theory.    
From another point of view,  if $K(z)$ is calculated analytically,  $h(z)$ must be  derived from a certain cosmological model (e.g. dark energy models in Friedmann-Robert-Walker (FRW) universe \cite{Ellis:2005wr,Pan:2015cqa}).  Meanwhile,  every value of $n$ is related to LIV model.     If we use both non-LIV assumption and LIV model  to constrain $E_{LV}$, it is  impossible to explain the results in theory.    

One solution of the non-LIV assumption  problem is  make one-to-one correspondence between  $n$ and  the related LIV gravity background. For example, when $n=1$, the calculation of $h(z)$ should be  based on Double Special Relativity.    This one-to-one calculation is restricted.  
In general,  the Lorentz variable $K(z)$ could be calculated  analytically and numerically.   Instead of calculating analytically, we   calculate $h(z)$  from  observational data.   In this section, we introduce the "cosmological model independent time delay method" which can avoid the non-LIV assumption.

 In observation, Planck Data favor $\Lambda$CDM cosmological model \cite{Ade:2015xua} which  has the fine tuning problem. Then, dynamical dark energy model and modified gravity are  used to explain the accelerating phenomenon  as well and they are indistinguishable by present  observational data. In view of degeneration,  we may regard the deviation between  cosmological model and  the LIV based  modified gravity as  unknown systematical errors.  Anyway,  using cosmological model  analytically  does not count systematical errors.  In this paper, we get the  $K(z)$ variable from the  observational data.     The  errors from observational data take the same role as the systematical errors in the accelerating  model degeneration.

\subsection{K Calculation}

As  no corresponding observations to $K(z)$ when  $n\neq0$, 
we use the technique of   Mean Value Theorem for Integrals to separate the observational and analytical parts in $K(z)$.   Assuming two nearby GRBs which  obey Eq.(\ref{testeq}),   a relative time-delay  could be gotten, 
\begin{eqnarray}
\label{tsub}
\Delta t_{obsh}-\Delta t_{obsl}= a_{LV}(K_h-K_l)+b_{sf}(z_h-z_l),
\end{eqnarray}
where the index $h$ ($l$) denotes high (low) redshift.
If   $D_h-D_l>0$   when $0< z_l < z_h$  and the function $1/h(z)$ does not change sign on the interval $[z_l, z_h]$,   Mean Value Theorem for Integrals  gives 
\begin{eqnarray}
\begin{split}\Delta K(z)=K_h-K_l& =\int_{z_l}^{z_h}\frac{(1+z)^{n}dz}{h(z)},\\
&= (1+z)^{n}|_{z_l\leq z\leq z_h}\int_{z_l}^{z_h}\frac{dz'}{h(z')}, \end{split}
\end{eqnarray}
where $z$ is a certain value in the range $[z_l, z_{h}]$. The choice of $z$ could be regarded as  a systematic error.
One simplistic way is to choose $z=(z_{l}+z_{h})/2$,  then
\begin{eqnarray}
\label{Ksub}
\Delta K (z)= (1+\frac{z_h+z_l}{2})^n (D(z_h)-D(z_l)).
\end{eqnarray}

After using Mean Value Theorem for Integrals, we could divide  the LIV effects into the $n$ part and the  dimensionless proper distance $D$.    If $K(z)$  is given by  cosmological model,  the unknown systematical errors between theories have been ignored.  In contrast,  
using observational  data to calculated $D$  take  all the errors in considerations.
As   Mean Value Theorem for Integrals   brings  systematical error to $\Delta K$, 
we divide  the error of  $(K_h-K_l)$ as  the observational error and the systematical error
\begin{eqnarray}
\label{Kerror}
 \sigma(\Delta K) & =\sqrt{ \sigma_{obs}^{2}+\sigma_{sys}^{2}},
\end{eqnarray}
where $ \sigma_{obs}=(1+(z_l+z_h)/2)^n \sqrt{\sigma_{D(z_h)}^{2}+\sigma_{D(z_l)}^{2} }$
and $\sigma_{sys}=n(1+(z_l+z_h/2)^{n-1} (z_h-z_l)(D(z_h)-D(z_l))/2$. 
 In this way, the error  is clear for the whole calculation of $K(z)$.  The most important improvement  of our CMITD method is  the correction of  
 errors.    After considering errors in  Eq.(\ref{Kerror}),   the calculation is consistent with theories.

\subsection{Regression}
Meanwhile,  we  derive a linear form for LIV effect based on  Eq.(\ref{tsub}):
\begin{eqnarray}
\label{lineregression}
\frac{\Delta t_{obsh}- \Delta t_{obsl}}{  z_h-z_l}= a_{LV}\frac{ K_h-K_l}{ z_h-z_l}+b_{sf}.
\end{eqnarray}
By defining $X=(K_h-K_l)/(  z_h-z_l)$ and
$Y=(\Delta t_{obsh}- \Delta t_{obsl})/(z_h-z_l)$, it becomes
\begin{equation}
\label{regression}
Y = a_{LV}X + b_{sf}.
\end{equation}
This is  a linear regression problem with  errors on both $X$ and $Y$.   $Y$ needs the time delay data and $X$ need  the distance data.  
  The Deming regression procedure provides such an unbiased estimation of  slope and intercept \cite{deming,linnet}. And,
 PyMC  is a python module that implements Bayesian statistical models, fitting algorithms and Markov chain Monte Carlo \cite{pymc}.  We combine  PyMC and Deming regression to do  Bayesian linear regression for LIV test.

\subsection{Five different Data Sets}
In practical test, we should put data into the CMITD method. For $X$ (or $K$), we use the GRB luminosity distance data where $D_{L}=D(z)/(H_0(1+z))$. 
The GRB luminosity distance dataset is based on Pad{\'e} method and Amati relation \cite{Liu:2014vda}.  This sample consists  $138$ long Swift GRBs  with redshift range $0<z<8.1$.   Its high-redshift ($z>1.4$) data are calibrated by the low redshift data. The low redshift are calibrated by    Union2.1 Data.     For $Y$ (or $t_{obs}$), we use the GRB time delay data.   The true time delay data is from Ref. \cite{Ellis:2005wr} which contains $35$ GRBs with a redshift range of   $0<z<6.29$. 

The luminosity distance data and the time delay data   have redshift matching problem.  For our LIV test,  based on the GRB luminosity data and Gaussian distribution, we   simulate  both non-LIV (Flat Simulated Data) and LIV  (Simulated  LIV Data)  time delay data. 
For Flat Simulated Data,  the  priors are set to $a_{LV}=0\pm0.001$, $b_{sf}=0\pm0.001$.
 For  Simulated  LIV Data,   the priors  are set to $a_{LV}=0.001\pm0.0001$, $b_{sf}=0.001\pm0.0001$.  The purpose of the two simulations is to  test the validity of the CMITD method.
 Then  the simulated priors  of $a_{LV}$ and $b_{sf}$ are assumed as  small ones.

\begin{figure}  \centering
      {\includegraphics[width=3.0in]{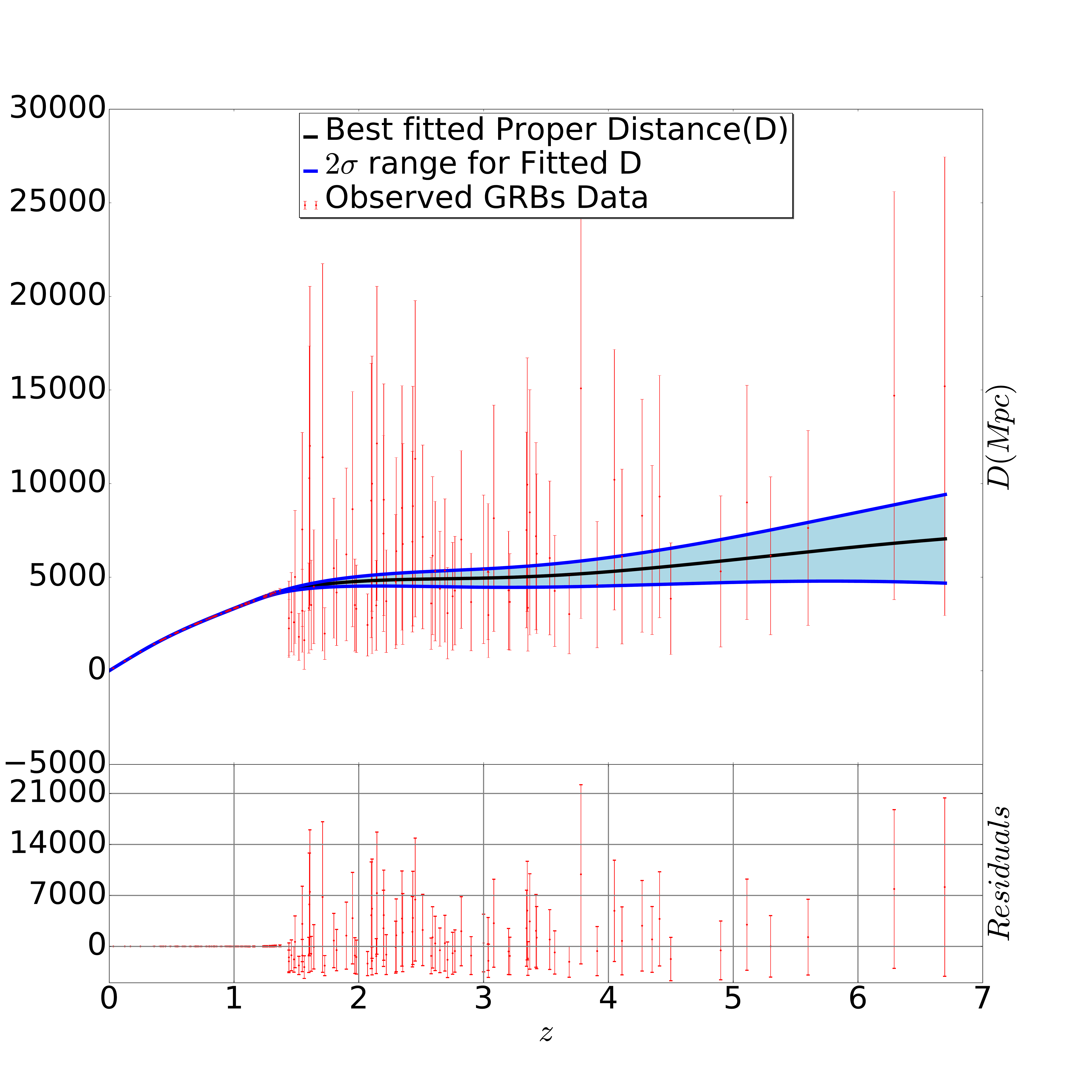}}\quad
                       \caption{ The red data points on the top are the proper Distance $D$ derived from GRBs luminosity distance data with its residual in the below panel. The blue region  are the $95\%$ C.L. range of GP and  the red line is the best fitted line of GP. }
  \label{dlresidual}
\end{figure}

We use GAPP  which is a Gaussian Process (GP) module written by python \cite{Seikel:2012uu} to do the non-linear regression for proper distance. And  we pick out the one have the same redshift with the  true time delay data.  We  tried to do GP on the true time delay as well.  As   the number of  time delay data is limited,  its best-fitted line is zero. The GP on time delay data gives a too strong prior. Therefore, we only do GP on the luminosity data. To supplement  the details, we plot the $D$ residuals in Figure \ref{dlresidual} which give  an increasing curve of  $D$ and consist with expanding universe scenario.    Increasing $D$ satisfies the condition of Mean Value Theorem of Integrals.

 As the luminosity distance  data  in the range  $z<1.4$  are calibrated by the Union2.1 data  \cite{Liu:2014vda},   the error of  low redshift $D$      are much smaller  than that of the  high redshift ones  which are calibrated from the  $z<1.4$ redshift data \cite{Liu:2014vda}.    To see the effect of  different  errors,   we choose  the data  in  $z\leq6.29$ range  as All Data and the data in $z\leq1.4$ range as Low Redshift Data. And to search the error effect of $X$, we remove the $X$ error by hand and call this dataset as  "$X$ error Removed Data". We summarize  the  five kinds of data in Table \ref{dataset}.

 \begin{table*}[t]
 \small
\begin{center}
\renewcommand\arraystretch{1.5}
	\begin{tabular}{|c|c|c|c|}
		\hline Data Sets  &  $X$ &  $Y$ & Prior  \\
		 \hline Flat Simulated Data    & True &  Simulated & $a_{LV}=0\pm0.001$, $b_{sf}=0\pm0.001$ \\
		 Simulated LIV Data    & True &  Simulated & $a_{LV}=0.001\pm0.0001$, $b_{sf}=0.001\pm0.0001$  \\	
		 	 All Data   & GP&   True & $-$  \\	  
		 Low Redshift Data   & GP&   True & $z<1.4$  \\	
			  $X$ Error Removed Data  & GP&   True &  Remove X error by hand  \\
		
			\hline
					\end{tabular}
\end{center}
\caption[crit]{ The five different Data Sets. GP and Simulation make the redshifts of X (luminosity distance data) and Y (time delay data) matched. }
\label{dataset}
\end{table*}

\begin{table*}[t]
\small
\begin{center}
\renewcommand\arraystretch{1.5}
	\begin{tabular}{|c|c|c|}
		\hline Models & Flat  Simulated Data& Simulated LIV Data   \\
		\hline $n=0.5$    &   $ a_{LV}= 0.000007 _{-0.000027-0.000055}^{+0.000027+ 0.000058}$& $ a_{LV}=   0.00022 _{-0.00009- 0.00018}^{+0. 00009+   0.00017}$ \\
		    &   $  b_{sf}= 0.00017_{-0.00017-0.00033}^{+0.00017+ 0.00033}$ &$ b_{sf}= 0.01089_{-0.00009- 0.00014}^{+0. 00009+   0.00016 }$  \\
		    
		\hline $n=1$& $ a_{LV}=  0.000007 _{-0.000028-0.000054}^{+0.000025+  0.000053}$   &$ a_{LV}=   0.00018 _{-0.00008-0.00017}^{+0. 00009+   0.00017}$   \\
		      &$ b_{sf}= 0.00023 _{-0.00020-0.00039}^{+0.00020+0.00039 }$ &$b_{sf}=  0.01107 _{-0. 00011-0.00021}^{+0. 0.00011+   0.00021  }$ \\	
		      
		      	\hline $n=2$   &$ a_{LV}=  0.000003 _{-0.000029-0.000054}^{+0. 000026+   0.000054}$ & $a_{LV}=0.00025 _{-0.00008- 0.00017}^{+0. 00008+   0.00016}$ \\	
		 &$ b_{sf}=  0.00052 _{-0.00030-0.00058}^{+0. 00030+   0.00058  }$ &$ b_{sf}= 0.01130 _{-0. 00014-0.00028}^{+0. 00014+   0.00028 }$    \\		
		\hline
		\end{tabular}
\end{center}
\caption[crit]{  Mean and $2\sigma$ values for $a_{LV}$ and $b_{sf}$ for Flat Simulated Data and  Simulated LIV data.}
\label{sim}
\end{table*}


\begin{table*}[t]
\small
\begin{center}
\renewcommand\arraystretch{1.5}
	\begin{tabular}{|c|c|c|c|}
		\hline Models   &  All  Data  &Low Redshfit Data &   $X$ error Removed Data \\
		\hline $n=0.5$    & $a_{LV}=    0.0548 _{-0.0681-0.1531}^{+0. 0757+ 0.1611}$  &   $ a_{LV}= -5.6142 _{-6. 4311-12.5395}^{+5. 6421+ 10.7426}$& $ a_{LV}=   0.2609 _{-0.1468- 0.2887}^{+0. 1466+   0.2889}$ \\
		     &$   b_{sf}=0.0332 _{-0.0504-0.1122}^{+0. 0485+ 0.1055  }$&   $  b_{sf}= 5.2039 _{-4.9938-9.4253}^{+5. 8457+ 11.3596  }$ &$ b_{sf}= -0.0714 _{-0. 0750- 0.1477}^{+0. 0752+   0.1481  }$ \\
		\hline $n=1$    &$a_{LV}= 0.0154 _{-0.1055-0.1479}^{+0. 0557+   0.1260}$ & $ a_{LV}=   6.6345 _{-13.2444-19.4092}^{+9.4344+  12.4488}$ &$ a_{LV}=   0.0978 _{-0.1108-0.2167}^{+0. 1106+   0.2196}$  \\
		       &$   b_{sf}=0.1148 _{-0.0650- 0.1536}^{+0. 1050+   0.1958  }$ &$ b_{sf}= -6.5102 _{-10.4932-13.4898}^{+14.8697+  21.4869  }$ &$b_{sf}=  -0.0033 _{-0. 1035-0.2045}^{+0. 1035+   0.2035  }$\\		\hline $n=2$   &  $  a_{LV}= 0.0025 _{-0.0168-0.0273}^{+0. 0115+   0.0247}$   &$ a_{LV}=  0.0616 _{-0. 3350-0.8864}^{+0. 4237+   1.0175}$ & $a_{LV}=0.0187 _{-0.0170- 0.0335}^{+0. 0170+   0.0333}$ \\	
		   &$ b_{sf}= -0.0310 _{-0. 0554-0.1169}^{+0. 0604+   0.1252  }$  &$ b_{sf}=  0.0678 _{-1.0007-2.1588}^{+0. 8382+   2.0100  }$    &$ b_{sf}= -0.1069 _{-0. 0775-0.1543}^{+0. 0776+   0.1515  }$ \\		
		\hline
		\end{tabular}
\end{center}
\caption[crit]{  Mean and $2\sigma$ values for $a_{LV}$ and $b_{sf}$ for All Data, Low Redshift Data and  $X$ Error  Removed Data.}
\label{true}
\end{table*}

 \begin{figure}  \centering
  {\includegraphics[width=2.7in]{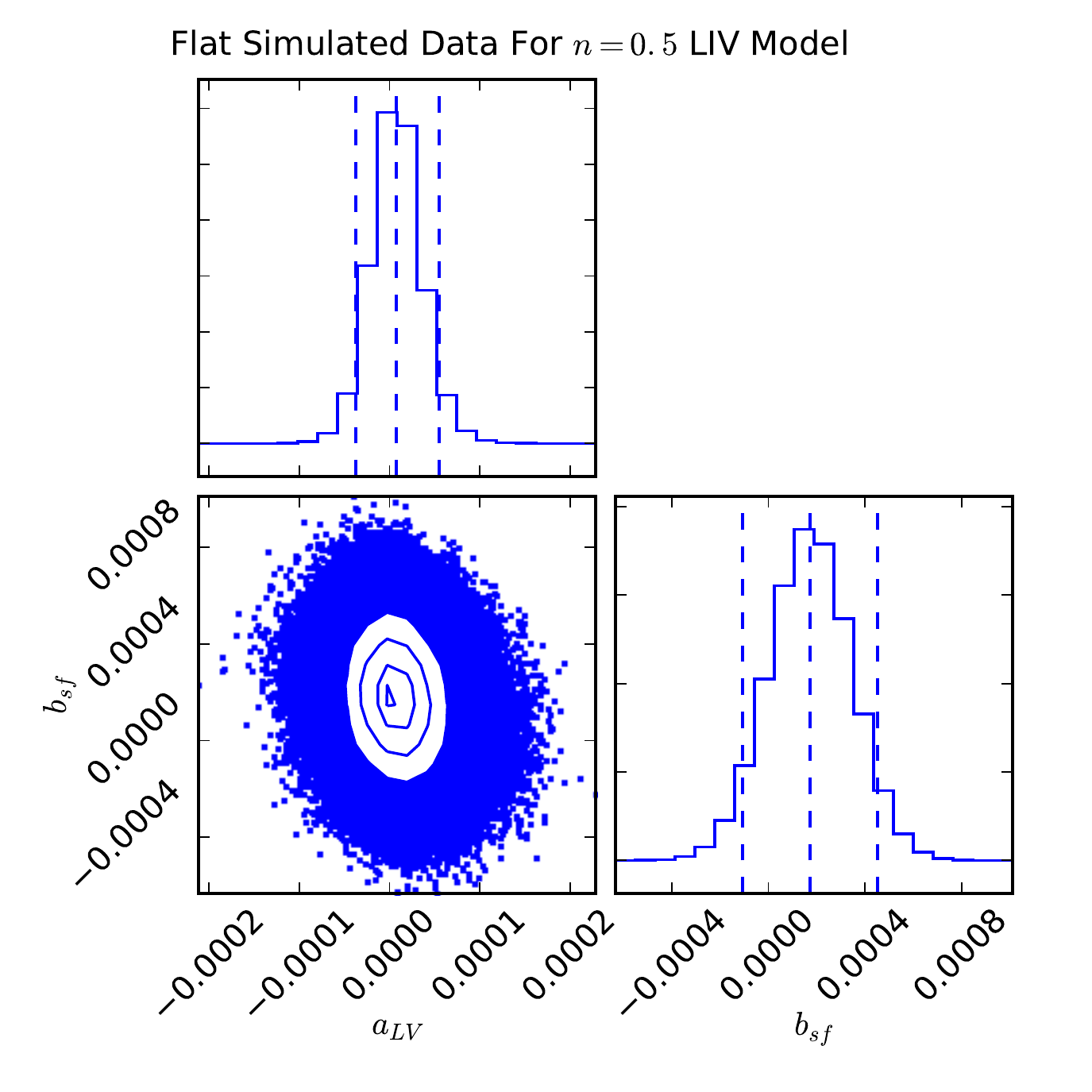}}
{\includegraphics[width=2.7in]{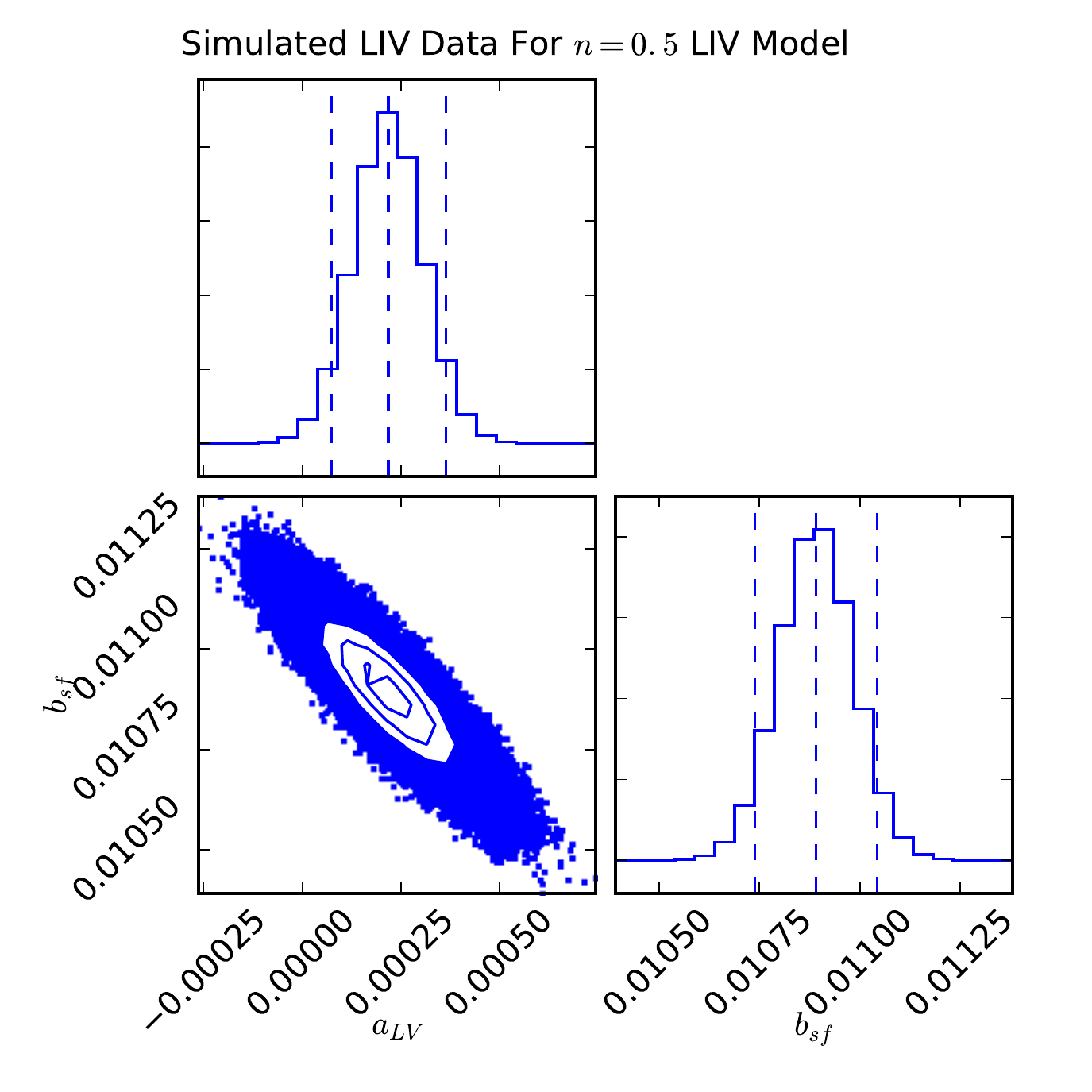}} \quad
 {\includegraphics[width=2.7in]{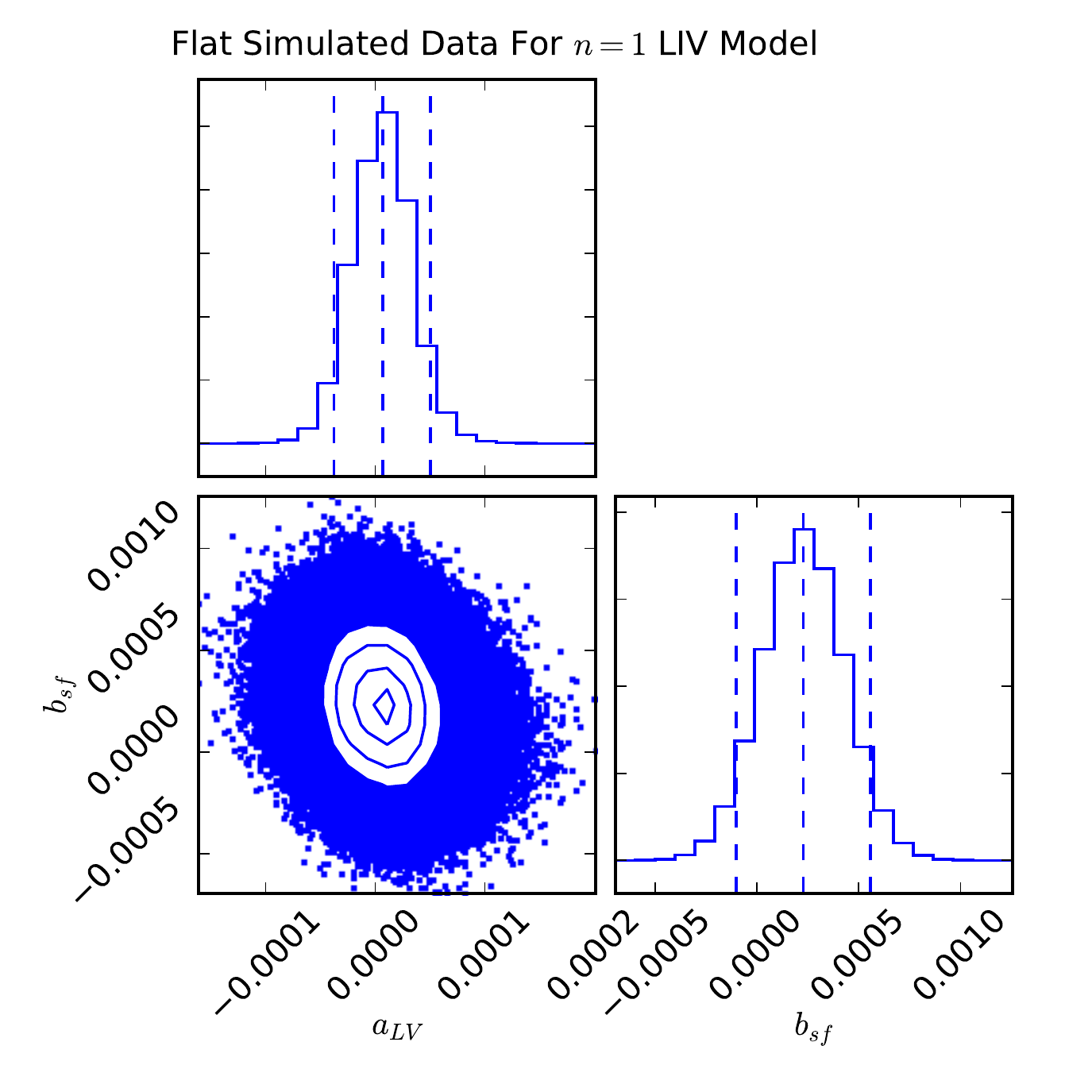}}
{\includegraphics[width=2.7in]{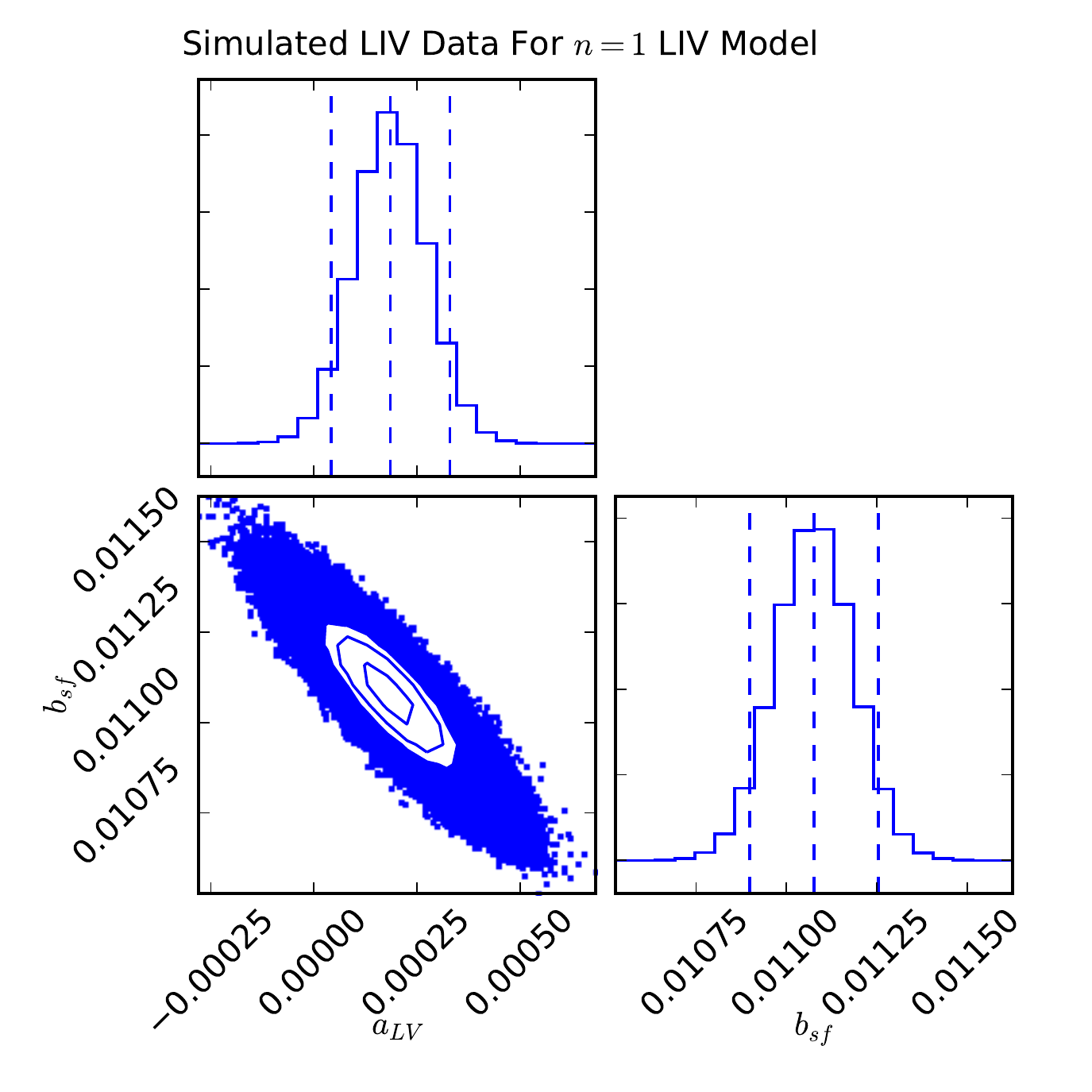}} \quad
  {\includegraphics[width=2.7in]{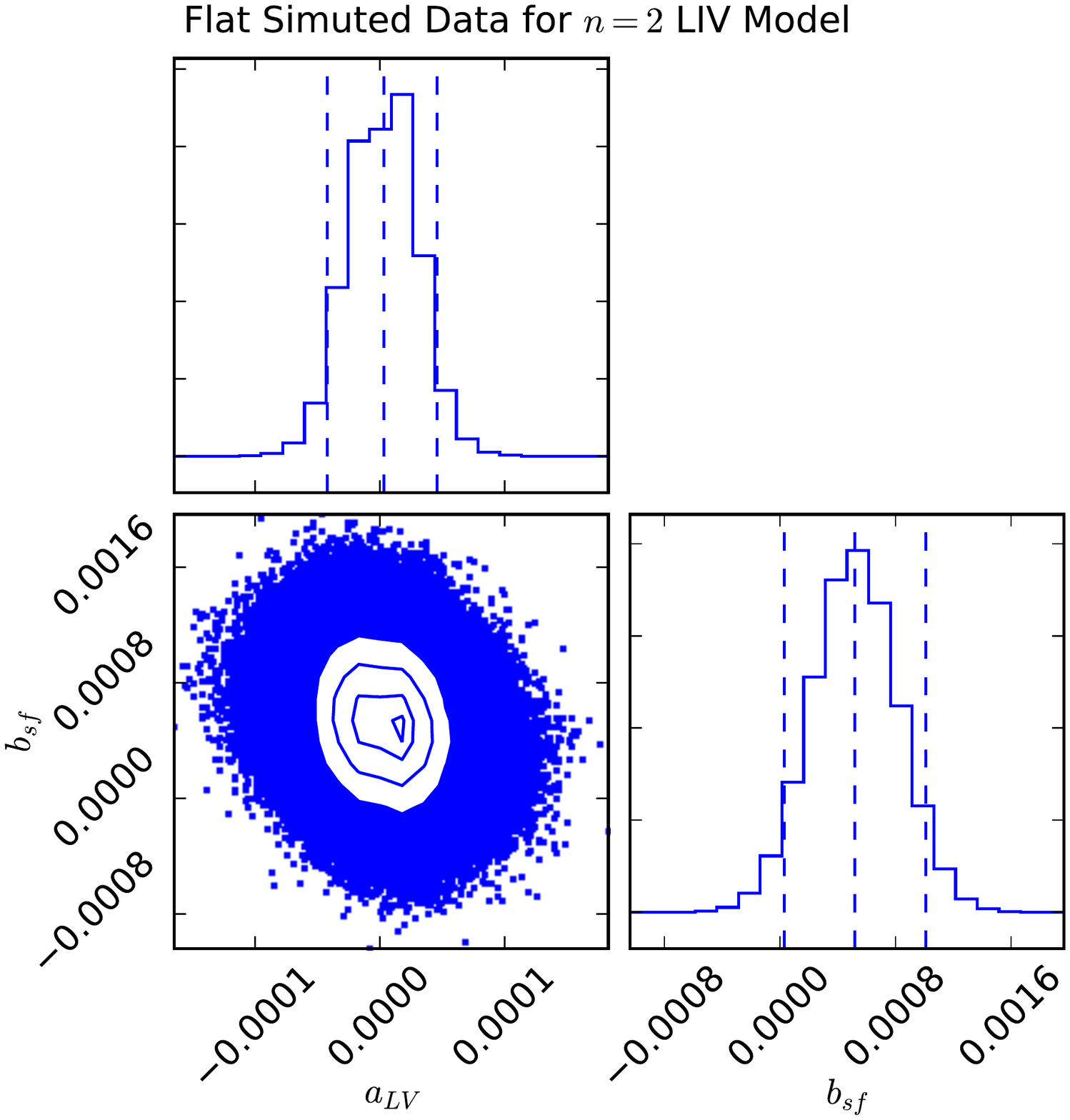}}
{\includegraphics[width=2.7in]{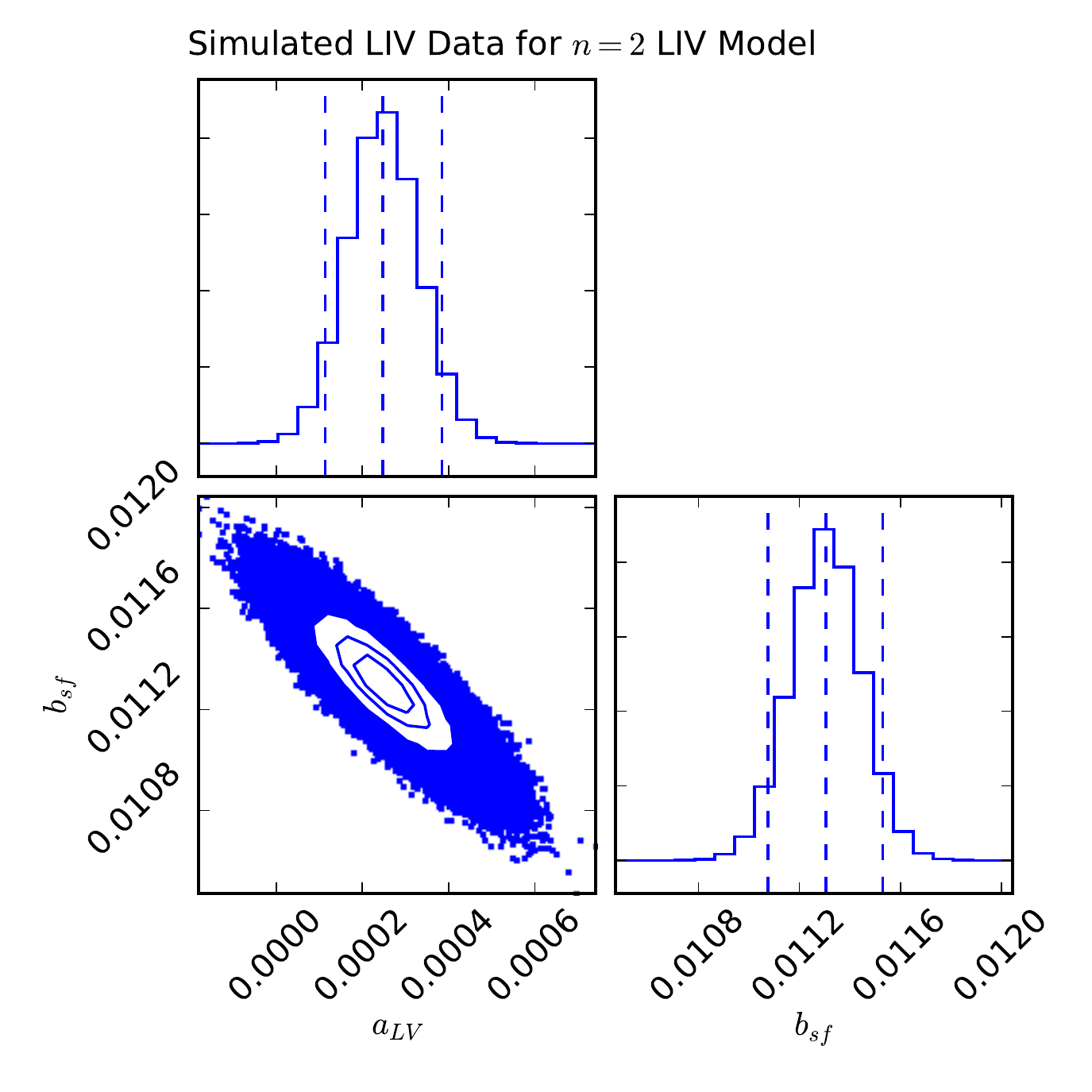}} \quad

        \caption{  The constraining results of $n=0.5$ LIV model.  Probability distributions  with the $1\sigma$ to $2\sigma$ contours corresponding to the parameter $a_{LV}$ and $b_{sf}$ are presented. The  panels are for Flat Simulated Data and Simulated LIV Data. }
  \label{figsim}
\end{figure}

 \begin{figure}  \centering
                   {\includegraphics[width=2.7in]{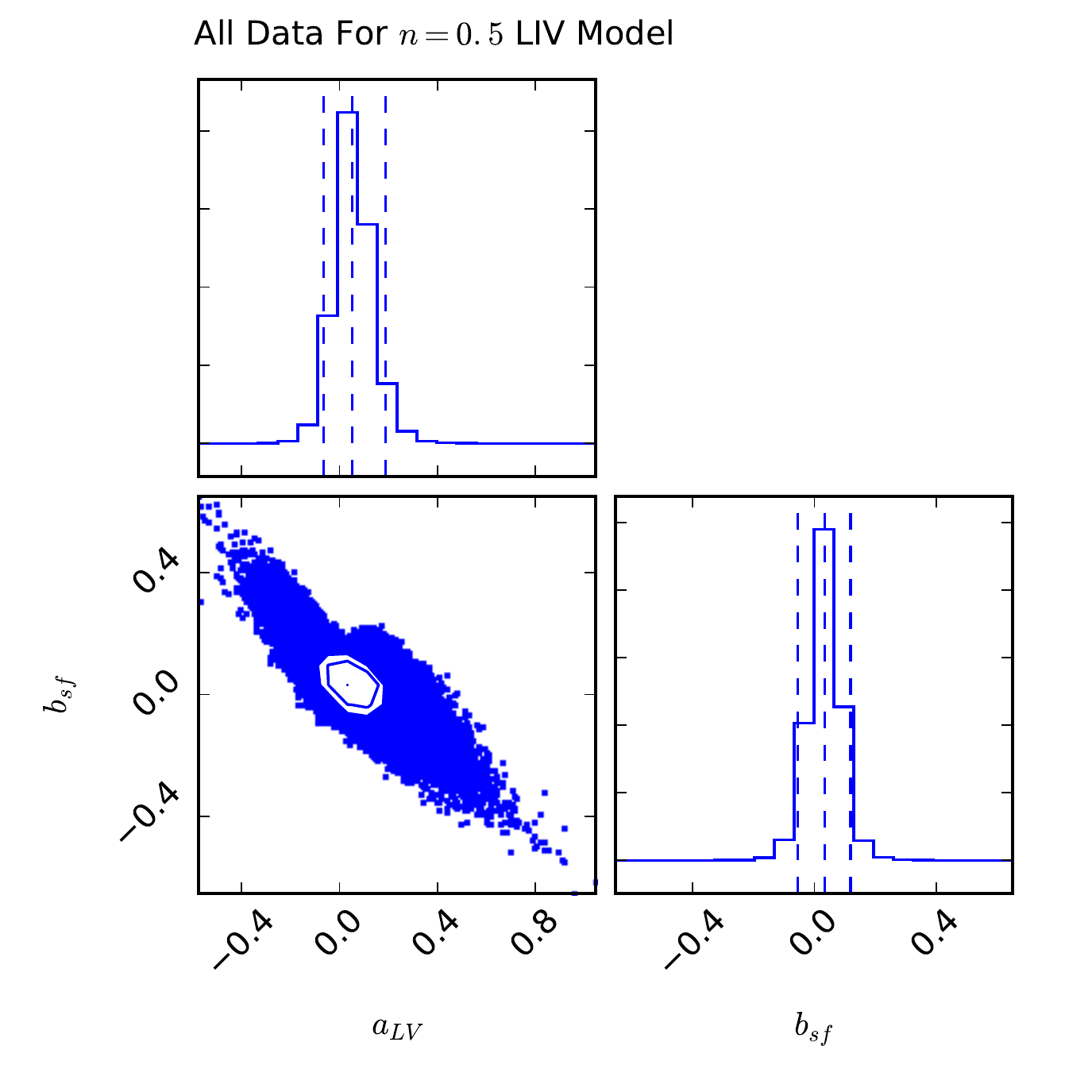}}
                   {\includegraphics[width=2.7in]{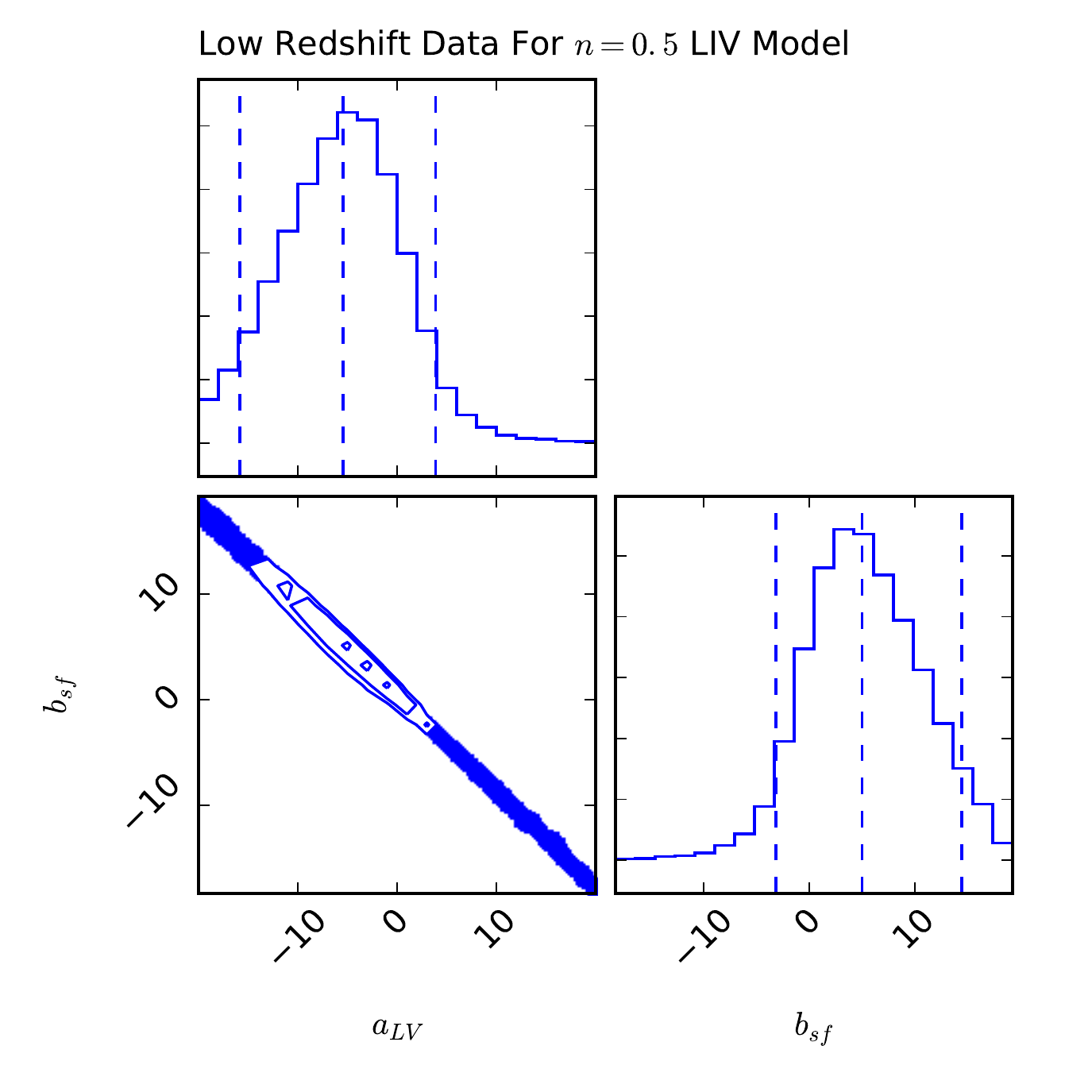} }\quad
                   {\includegraphics[width=2.7in]{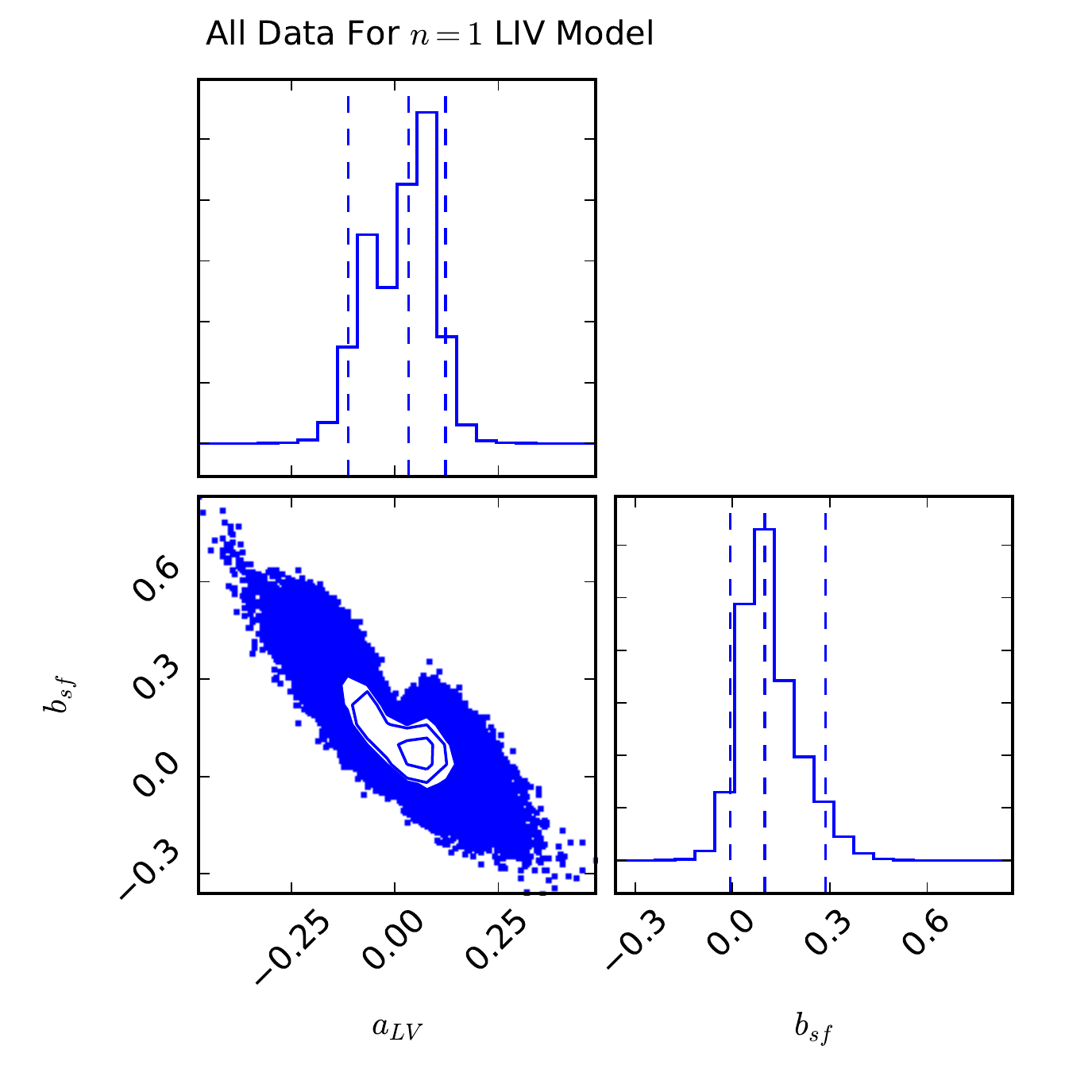}}                   
                                           {\includegraphics[width=2.7in]{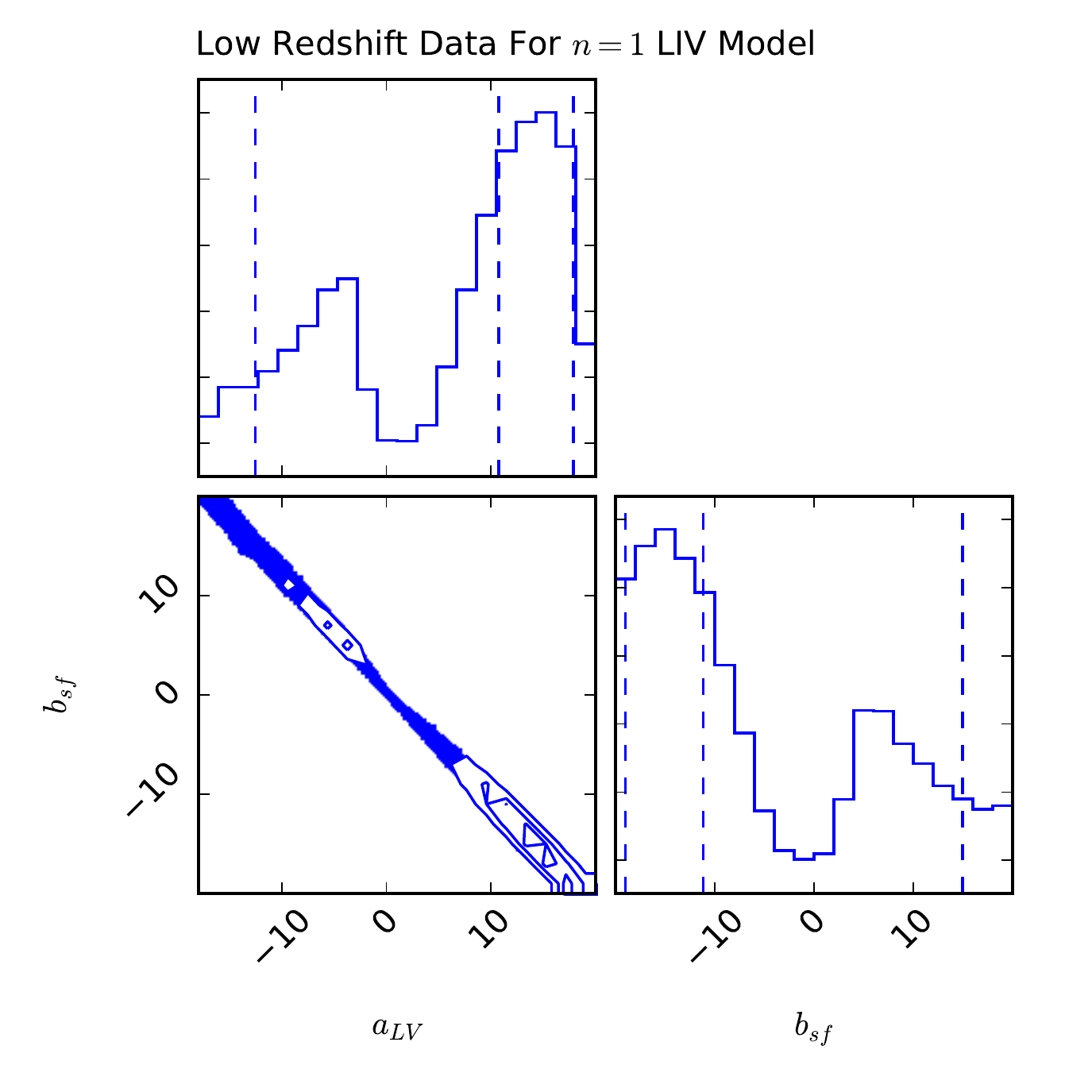}} \quad
                                             {\includegraphics[width=2.7in]{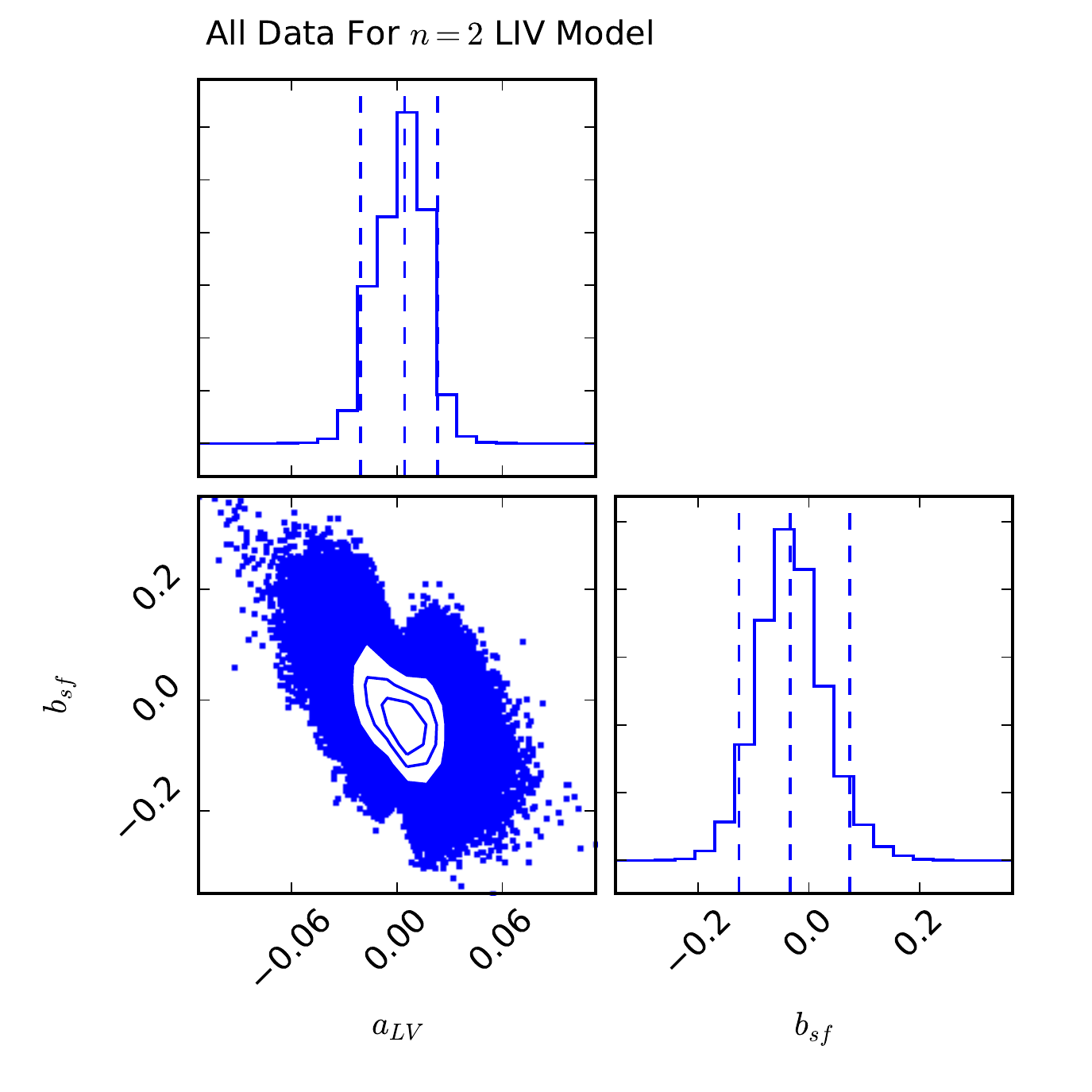}}
                                                        {\includegraphics[width=2.7in]{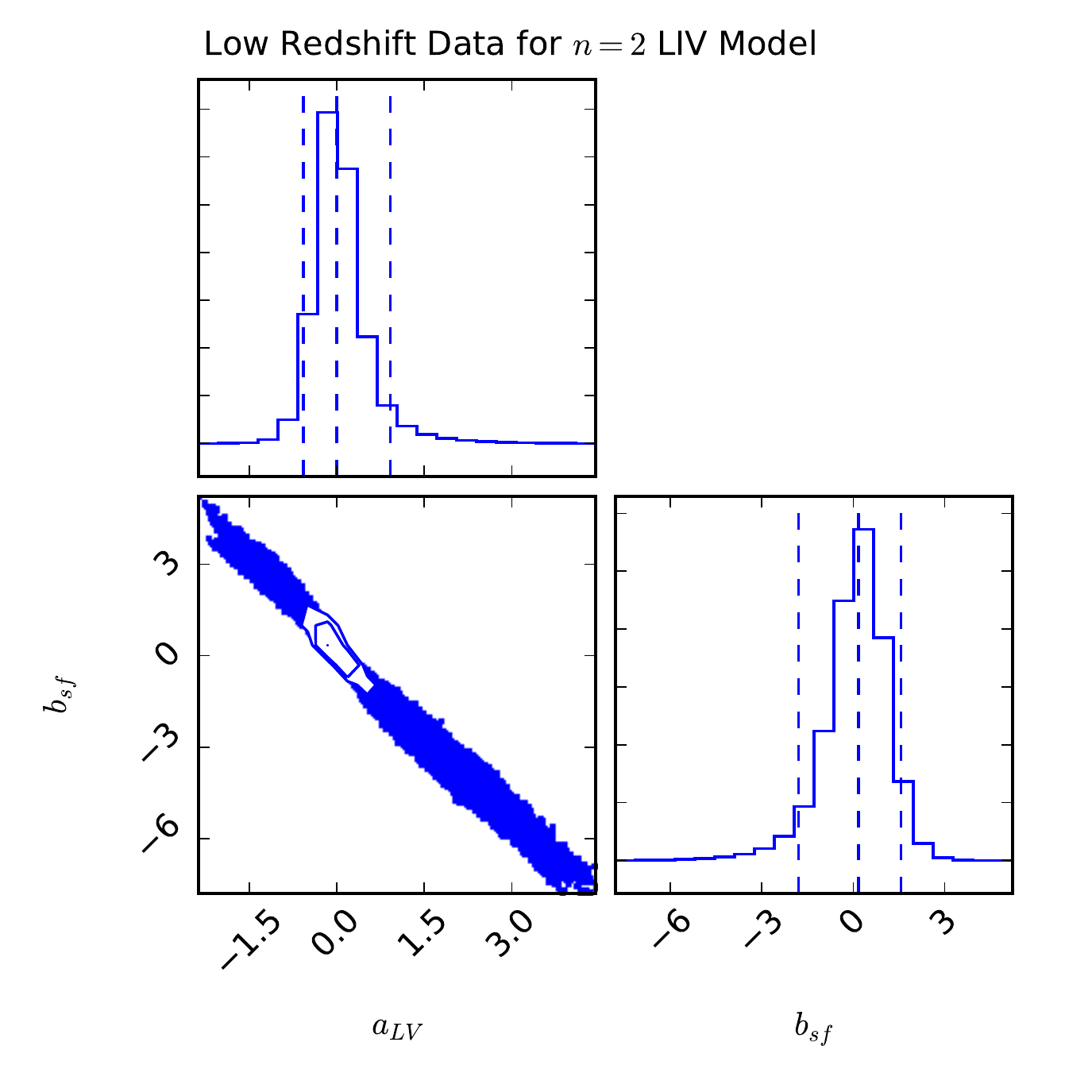}}\quad
 \caption{ Same as Figure \ref{figsim}, but for  All Data and Low Redshift Data  separately.  }
  \label{figtrue}
\end{figure}



 \section{ Results and Discussion}\label{sec4}
   Based on our five data sets, we calculate the relative value of $K$ by  Mean Value Theorem for Integrals and   fix  $H_0=67.8 km/s/Mpc$  following Planck experiment's suggestion  \cite{Ade:2015xua}.   PyMC and Deming regression are used to get the propagation and intrinsic value of $a_{LV}$ and $b_{sf}$. We run the MCMC chains  $20000000$ steps and present the marginalized distribution for parameters which is a 2-dimensional histogram and list $1\sigma$ and $2\sigma$ constraining ranges of $a_{LV}$ and $b_{sf}$ in Tables \ref{sim} and \ref{true}.   Our  corner   plots  \cite{corner} in Figures \ref{figsim} and \ref{figtrue}  show  the marginalized  distributions of $a_{LV}$ and $b_{sf}$.


 \subsection{Results from the Simulated Data}
  
   Once  defining  the measurement errors correctly, the Deming regression procedure provides an unbiased estimate of the slope \cite{linnet}.   As  Figure \ref{figsim} and Table \ref{sim} show,  the results of  Flat Simulated Data present that $a_{LV}=0$ and $b_{sf}=0$ are in  $1\sigma$ range. And, through our prior precision is assumed as $0.001$,  the results reach to $0.00001$ for $a_{LV}$ and $0.0001$ for $b_{sf}$ which is much tighter. 
  For the Simulated LIV Data,  the LIV slope $a_{LV}$ and the intrinsic time-lag $b_{sf}$ are  negative correlated. And $a_{LV}\neq0$ and $b_{sf}\neq0$ are in $2\sigma$ range.  Through our prior precision is assumed as $0.0001$, its results reach to $0.0001$ for $a_{LV}$ and $0.0001$ for $b_{sf}$. The results is tighter than the prior of $a_{LV}$ and looser than that of $b_{sf}$.  Both results from the simulated  data are consistent  with the priors.

  The  parameters   $a_{LV}$ and $b_{sf}$ are the propagation and intrinsic effects, their posteriors should be Gaussian  distribution.     And, regression models work well with symmetrical and bell-shaped curves.  
 As  Figure \ref{figsim} and Table \ref{sim} show, the  marginalized distributions in simulated data are symmetrical  bell-shaped curve. The results consist  with the Gaussian distribution theory assumption.    By comparing the priors and results for  the two simulated data,  the conclusion is  that  we could get  the correct LIV results by our  effective CMITD method.

\subsection{Results form the True Data}
As Figure \ref{figtrue}  and Table \ref{true} show,  All  Data give  symmetrical and bell-shaped curves for  $n=0.5$,$n=1$,$n=2$ LIV  models. The luminosity and time delay data from GRBs could separate the propagation effect  $a_{LV}$  from the intrinsic source effect $b_{sf}$.
 The best fitted  $a_{LV}$  and $b_{sf}$  are around $0$. Especially, the $1\sigma$ ranges contain the Lorentz invariance value $a_{LV}=0$.

And, the Low Redshift Data do not present good results of  LIV  because the marginalized distributions of parameters are not symmetrical  bell-shaped curves.   The $1\sigma$  (or $2\sigma$) ranges  of Low Redshift Data cover the $1\sigma$  (or  $2\sigma$) ranges  of All Data whose results are   preciser. Moreover, the $n=1$ model whose results has two peaks does not consist with the Low Redshift Data.  As All Data include not only the Low Redshift Data  but also high redshift data where $z>1.4$,  the CMITD method needs high redshift data.  Then, the  cosmological-distance-like term $K(z)$ become detectable by accumulating  the small LIV effect over  a long particle travel time.

  If we  remove the error of  $X$ by hand, 
the ranges of   $a_{LV}$ and $b_{sf}$ are  more symmetric but larger  than the ones from  All Data as Table \ref{true} shows.   However,  when $n=0.5$ and $n=2$,   $a_{LV}=0$ is not in the $1\sigma$ range for   $X$ Error   Removed Data while $a_{LV}=0$ is in $1\sigma$ range for All Data.  It  is the main difference between All Data  and  $X$ Error  Removed Data which means removing $X$ error by hand makes the constraining results biased.    As calculating  $K$ from cosmological model  is degenerated  to Removing  $X$ error by hand, the LIV test results derived from cosmological model  are biased as well.

  \begin{table*}[t]
  \small
\begin{center}
\renewcommand\arraystretch{1.5}
	\begin{tabular}{|c|c|c|c|}
		\hline $s_{\pm}$ & $n=0.5$ &  $n=1$ &  $n=2$\\
		\hline $+1$   &   $ E_{LV}\ge 8.00\times 10^{32}GeV$& $E_{LV}\ge 1.03 \times 10^{15} GeV $ &$E_{LV}\ge1.31 \times 10^{6}GeV $\\
(Subluminal) &   $1-v/c\leq4.74\times 10^{-19}  $& $1-v/c\leq  3.11\times 10^{-19} $ &$1- v/c\leq 5.97\times 10^{-20} $  \\
		\hline $-1$   &   $E_{LV} \ge3.85\times 10^{33}GeV  $& $ E_{LV}\ge  1.10\times 10^{15} GeV $ &$E_{LV}\ge  1.37 \times 10^{6}GeV $\\
		(Superluminal)      &   $1-v/c \ge-2.16\times 10^{-19} $& $  1-v/c\ge-2.91\times 10^{-19}$  &$ 1- v/c\ge-5.44 \times 10^{-20} $  \\		
		\hline
		\end{tabular}
\end{center}
\caption[crit]{  The range of $E_{LV}$ and velocity by considering $E_{h^{*}}=320keV$, $E_{l^{*}}=25keV$ \cite{Ellis:2005wr} and the lower limit of $95\%$ confedence level of $a_{LV}$.  }
\label{elvv}
\end{table*}  

  \subsection{Discussions}
  
Through our CMITD method is cosmological model independent, the observational data may be regarded as model dependent. 
  Considering  the comparison between  Low Redshift Data and All data  in the LIV test, we need high redshift data.
Therefore, the choices of data  for LIV test  are restricted.    
  
  Our  GRB luminosity data in LIV test are based on  Amati relation which is challenged by the  selection effect  \cite{Collazzi:2011rn}.  Indeed,  most experience relations are in debate, e.g. the Ghirlanda relation  has the redshift ambiguity problem at substantially higher redshift \cite{Collazzi:2011rn}.    Here,    the  GRB luminosity  data  are calibrated by    the Union 2.1 Data   which are related with $\Lambda$CDM model. Then,   it is not totally free from cosmological model. 
      But the Simulated Data which  use the true luminosity data  could distinguish LIV effect and   their constrained  $a_{LV}$ and $b_{sf}$ obey Gaussian distributions.  In a conclusion,   the effect caused by cosmological model in GRB luminosity data is small  and  the validity of our CMITD method is not affected.     In contrast,   the true time delay data is cosmological model dependent. Take the $z=0.45$ data for example \cite{Price:2002xs}, to estimate the properties of the host galaxy, the authors adopted a standard cosmology with fixed $H_0$ and  $\Omega_{m0}$.     As the time delay data related to the standard cosmology without  LIV effect, the test result does not show any LIV signals.      The results given by CMITD method  consist with data prior.    
  
  Through using dark energy model is  regarded  as  losing unknown errors   in LIV test,  the  comparison among   various constraining results is still interesting.   For $\Lambda$CDM model, Ref.\cite{Pan:2015cqa} gives
$a_{LV} = -0.017^{+0.0717+0.1416}_{ -0.0718-0.1415} $ and
$b_{sf} = -0.00013^{+0.0154+0.0308}_{ -0.0155-0.0305}$ by using the $\chi^{2}$ methods with the GRBs' time delay data, the cosmic microwave background data from the Planck first year release, the baryon acoustic oscillation data  and Union2 type Ia supernovae data.
  Our  results from CMITD method present $a_{LV}= 0.0154 _{-0.1055-0.1479}^{+0. 0557+ 0.1260}$  and   $b_{sf}=0.1148 _{-0.0650- 0.1536}^{+0. 1050+0.1958  }$ by only using the GRBs' time delay and luminosity distance data.   We get a comparable result with the $\Lambda$CDM case in Ref.\cite{ Pan:2015cqa} by using less observational data and no cosmological model.   The $1\sigma$ and $2\sigma$ ranges  of $a_{LV}$ are at the same order as  the results derived from$\Lambda$CDM cosmological model,  and the $1\sigma$ and $2\sigma$ ranges  of of  $b_{sf}$ are one order larger than that derived from $\Lambda$CDM  cosmological model.

And, we list  the LIV energy scale $E_{LV}$ in  Table \ref{elvv}.
 The  parameter $n$ affects the LIV energy scale heavily. The smaller $n$ is, the larger the LIV energy scale is. 
  The   $E_{LV}$ parameter  is  larger than  $10^{32} GeV$, $10^{15} GeV$ and $10^{6} GeV$ for $n=0.5$, $n=1$, $n=2$ separately.
The LIV energy scale of  $n=0.5$ model is much  larger than the Planck scale $E_{pl}=\sqrt{\hbar c^{5}/G}=1.2\times 10^{19}GeV$.
 As the choice of $n$ in Multifractal Spacetime could be in the range of $0<n<1$,
 if the energy scale of LIV was around Planck scale,  the Multifractal Spacetime Theory should choose $n>0.5$.
  As $E_{LV}$ is too small to be effective, the $n=2$ LIV model need more data break the  degeneration between $E_{LV}$ and $n$. 
The velocity constraints from photon $|1-v/c|$ are  at $10^{-19}-10^{-20}$ order. It is also possible that Lorentz violation  manifest itself in the gravitational sector.
  As the observation of the Gravitational Waves (GWs) from the neutron star binary coalescence GW170817 and of the associated Gamma-ray burst GRB 170817A \cite{Monitor:2017mdv} gives $-3\times 10^{-15}<v/v_{GW}-1< 7\times 10^{-16}$, the photon part are preciser than the present GWs observation.

\section{Conclusion}\label{sec5}

    General Relativity   and Lorentz invariance violation  are two contradictory theories.  As LIV effect depends on $n$ and dark energy model depends GR,  unknown systematic errors are given when using dark energy model  in LIV tests. 
In  this paper,  we proposed   a cosmological model independent time delay method to  test the Lorentz invariance violation. Five different kinds of Data combinations are used.  
 The simulated time delay data show the method is effective to detect LIV.
The true  time delay data  present  non-LIV results because of  the data assumption.   By comparing the results from Low Redshift Data and  All Data,  we conclude that  high redshift data are needed of LIV test.  By comparing the results from  All data and  $X$ Error Removed Data, we conclude that  the error of $X$ is critical for the existence and magnitude  of LIV. 
 If the future detections give out  model-independent time delay data,  more essence of physics could be extracted.



\section*{Acknowledgements}
We are  grateful  for the useful comments from Prof. Naqing Xie, Prof. Bin Hu and Prof. Hao Wei.
YZ is supported by
CQ CSTC under grant
No. cstc2015jcyjA00044 and  No. cstc2015jcyjA00013, CQ MEC under grant No. KJ1500414, and HZ is supported by National Natural Science Foundation of China under Grant Nos. 11075106, 11275128 and 11105004.



\end{document}